**Keynotes on membrane proteomics**

*Thierry Rabilloud*

CEA, DSV, DRDC, ICH, Laboratoire d'Immunochimie, 17 rue des martyrs,  F-38054 GRENOBLE CEDEX 9,  France; INSERM U548

**1 Introduction**

Before addressing the key issues pertaining to proteomics of membrane proteins, which is a key subject in the wider topic of subcellular proteomics, the first question that must be asked is the definition of a membrane protein. While this might seem trivial at the first glance, the answer is far from being obvious. A commonly-accepted definition for a membrane protein is a protein associated with a membrane, i.e. a lipid bilayer. However, the meaning of the word "associated" in this case is the subject of almost infinite debate. In fact, two different situations arise. In the first, simple situation, the polypeptide chain spans the lipid bilayer a certain number of times. These proteins are defined as integral or intrinsic membrane proteins, as their association with the lipid bilayer is doubtless.. The second, much more complicated case, is when the association with the lipd bilayer is not achieved by transmembarne segments or by transmembrane barrels (e.g. in porins), because many other types of association with membranes are encountered. In one particular type, the association is mediated by a post translational modification of the polypeptide, namely the grafting of an fatty acid or polyisoprenyl chain. Polyisoprenylation has been demonstrated to result in physical association with the membrane, as shown for small G proteins (Verma et al. 1994)  the result of a simple acylation is probably less straightforward. Strong association of proteins with lipid membranes is also achieved by glycolipid anchors such as the glycosylphosphatidyl inositol common in eukaryotic species, or, as in bacterial lipoproteins, by an *N*-acyl diglyceride linkage to an *N*-terminal cysteine residue that becomes available after cleavage of a signal sequence (Cross 1991).  Last but not least, the protein can also be acylated by a simple fatty acid chain. In this case, the reality of the anchoring to the lipid bilayer is much less clear.

Protein-membrane association via a post-translational modification introduces the notion of dynamic association and partitioning of proteins between the membrane phase of the cells and the aqueous phase (cytosol or inner phase of organelles). Consequently, such proteins can be found both as membrane-associated and membrane-free, which is not the case with intrinsic membrane proteins which are strictly membrane embedded. Another type of association to membrane is mediated by protein-protein interactions with other membrane proteins. A typical example of this situation is provided by the respiratory complexes. In the case of ubiquinol-cytochrome c oxidoreductase, core proteins 1 and 2 does not show any interaction with the lipid membrane, but only with the protein subunits spanning the membrane (e.g. cytochrome b) (Iwata et al. 1998).

However, the situation is often much less clear than this one, and drives quite fast into the infinite debate of what is a membrane protein apart from the integral ones. In addition to the dynamic view of association to membranes, this complicated situation arises mainly from the operational, biochemical, definition of membranes. When a cell is lysed in an aqueous, detergent-free, medium, the cell-limiting membrane and the network of inner membranes (when applicable) will fragment into vesicles which can be separated from the bulk of cytosol by sedimentation or partition techniques. These techniques are also able to separate membrane-bound organelles (e.g. mitochondria and plasts) from other cellular components, providing good purity. However, not every protein present in such preparations can be considered as a membrane protein. For example, when vesicles will form from larger structures upon lysis, simple physical entrapment will bring soluble proteins in the lumen of the vesicles. Furthermore, it appears more and more clearly that some classes of proteins (e.g. cytoskeletal proteins or ribosomal proteins) are directly or indirectly associated to bona fide transmembrane proteins or to the lipid bilayer. Should these proteins be classified as membrane proteins ?   Should the non-membrane spanning subunits of respiratory complexes be considered as membrane proteins ? This gives an example of the debate that can take place on the notion of membrane proteins.

This could be seen as a very theoretical debate, but this has indeed very practical implications. Let us take the example of the analysis of a reticulum preparation. Endoplasmic reticulum is the place of synthesis of most secreted proteins and of most transmembrane proteins. As such, it contains many ribosomes (the Rough ER) and the reticulum vesicles are known to be associated with cytoskeletal filaments. The problem arises from the fact that the protein content of these- cytoskeletal and ribosomal « contaminants » is concentrated in a few proteins. This means in turn that even if these « contaminants » represent a few percent of the total protein mass of the preparation, the few proteins represented will be easily detected by any proteomics analysis method.
Conversely, the « real » reticulum proteins are scattered among almost all the secreted proteins and  transmembrane proteins of the cells, plus the resident proteins of the ER. This means in turn that these proteins will appear as less prominent upon proteomics analysis, just because each protien species is more diluted than the few « contaminating » ones.

## 2. Sample preparation issues for membrane proteomics

A consequence of the above considerations is that sample preparation is very critical in membrane proteomics. As the volume occupied by the lipid bilayer in a membrane sample is very small in comparison to the volume of the aqueous phase, the transmembrane proteins must be considered as rare components of the sample, although they often drive most of the interest of the researchers. In addition to that, these transmembrane proteins often pose a very difficult solubilization problem. Transmembrane proteins usually contain domains protruding in the extracellular environment or in the cytosol. These domains behave as classical protein domains and are therefore stable in a water-based solvent. However, membrane proteins also contain domains that are embedded more or less deep in the lipid bilayer. Consequently, these domains are stable in a hydrocarbon-like environment. If a membrane protein is to be solubilized prior to its purification, this means that the solvent which is used to this purpose must show at the same time water-like and hydrocarbon-like properties. A water-based solvent will induce aggregation of the membrane domains via hydrophobic interactions, and thus protein precipitation. However, an organic solvent will induce in most cases precipitation of the membrane proteins via their water-soluble domains, which are denatured and coalesce in such a solvent. Some very hydrophobic proteins, however, are soluble in organic solvents (e.g. in (Molloy et al. 1999), (Blonder et al. 2003) ). This is due to the reduced size of their non-membrane domains, and to the fact that all water-soluble protein domains contain a hydrophobic core, which can be soluble in organic solvents. In some cases, these positive solubilization forces can overcome the precipitation-deriving forces and make the protein soluble in organic solvents.

This is however not a general case, and the general rule for membrane proteins is that they require a solvent that has as the same time strong water-like and strong hydrocarbon-like properties. This is not the situation of mixed solvents, i.e. mixtures of water and water-miscible solvents, which only offer average properties and not a combination of both properties. Such a combination is only offered by a stable dispersion of hydrocarbon chains in a water-based solvent. This is the definition of lipid membranes, but such assemblies are very difficult to handle along a purification process. Hopefully, this definition is also the one of detergent micelles, which are much easier to handle along a purification process. These constraints explain why detergents are some kind of a universal tool for the disruption of cell membranes and for the solublization of membrane proteins.

The micelle-forming ability of detergents is driven by their chemical structure, which combine a hydrophobic part (the tail), promoting aggregation of the molecules into the micelles, which is linked to a hydrophilic one (the head) promoting water solubility of the individual molecules and of the micelles. By using different structures or heads and tails and by combining various tails with various heads, an almost infinite range of detergents can be generated. Their protein and lipid solubilization properties are of course strongly dependent on the chemical properties of the heads and tails. As a rule of thumb, rigid tails (e.g. steroid-based) and weakly polar heads (e.g. glycosides, oligoethylene glycol) generally lead to "mild" detergents, i.e. chemicals that have good lipid solubilization properties, but that have weak protein dissociation properties. While these detergents do not denature proteins, they can prove unable to prevent protein-protein interactions promoting precipitation. Consequently, their protein solubilization properties are highly variable, and a lengthy optimization process in the detergent choice is usually needed for various membrane proteins.

Conversely, detergent having a flexible tail and a strongly polar (i.e. ionic) head are viewed as "strong" detergents. In addition of their lipid solubilizing properties, these detergents are able to disrupt the hydrophobic interactions maintaining the structure of the proteins. They are therefore denaturing. However, because of their ionic nature, the bound detergent imparts a net electrical charge to the denatured proteins and induces a strong electrostatic repulsion between protein molecules. Thus, even denatured proteins can no longer aggregate, and these ionic detergents are very powerful protein solubilizing agents.

These protein solubilization conditions have a key impact on the protein fractionation that can be carried out afterwards, in the sense that they will restrict the choice to techniques with which they are compatible. As an example, ionic detergents are not compatible with any technique using protein charge or pI as the fractionation parameter.

All of the above is mainly true when some fractionation is to be carried out at the protein level. In some approaches, the fractionation is carried out only on peptides arising from the digestion of the membrane preparation. In this case too, the choice of the solubilization media will be dictated by the constraints imposed by the subsequent peptide fractionation process. However, the use of chromatographic peptide fractionation tools, and especially those based on reverse-phase chromatography, will make the use of detergents more problematic, leading to specially-designed, detergent-free protocols (Wu et al. 2003, Fischer et al. 2006).

**3. Issues linked to the separation process**

Once the membrane proteins have been solubilized, usually in a water-detergent medium, they must be fractionated. As each fractionation method brings its own constraints, this will further restrict the scope of possibilities that can be used for sample solubilization. The following sections will exemplify some constraints linked to the most commonly used fractionation processes in proteomics

## 3.1. Constraints in proteomics based on zone electrophoresis

In these proteomics methods, the separation process is split in two phases (e.g. in Bell et al. 2001). The first phase is a protein separation by denaturing zone electrophoresis, i.e. in the presence of denaturing detergents, most often sodium dodecyl sulfate (SDS). The second phase is carried out by chromatography on the peptides produced by digestion of the separated proteins. This has no impact on the sample preparation itself, which just needs to be compatible with the initial zone electrophoresis.

This is by far the simplest case. Sample preparation is achieved by mixing the initial sample with a buffered, concentrated solution of an ionic detergent, usually containing a reducer to break disulfide bridges and sometimes an additional nonionic chaotrope such as urea. Ionic detergents are among the most powerful protein denaturing solubilizing agents. Their strong binding to proteins make all proteins to bear an electric charge of the same type, whatever their initial charge may be. This induces in turn a strong electrostatic repulsion between protein molecules, and thus maximal solubility. The system of choice is based on SDS, as this detergent binds rather uniformly to proteins. However, SDS alone at room temperature, even at high concentrations, may not be powerful enough to denature all proteins. This is why heating of the sample in the presence of SDS is usually recommended. The additional denaturation brought by heat synergizes with SDS to produce maximal solubilization and denaturation. However, some hydrophobic transmembrane proteins do not withstand this heating step. In this case, their hydrophobic parts coagulate through the effect of heat much faster than binding of SDS can solubilize them. This leads to precipitation of these proteins.

The use of SDS is not always without drawbacks. One of the most important is encountered when the sample is rich in DNA. A terrible viscosity results, which can hamper the electrophoresis process. Moreover, some protein classes (e.g. glycoproteins) bind SDS poorly and are thus poorly separated in the subsequent electrophoresis. In such cases, it is advisable to use cationic detergents. They are usually less potent than SDS, so that a urea-detergent mixture must be used for optimal solubilization (MacFarlane 1989). Moreover, electrophoresis in the presence of cationic detergents must be carried out at a very acidic pH, which is not technically simple but still feasible (MacFarlane 1989). This technique has however gained recent popularity as a double zone electrophoresis method able to separate even membrane proteins (Hartinger et al. 1996), and showing more separation power than SDS electrophoresis alone.

An important and often overlooked variegation of zone electrophoresis-based separations is the mixed native-denaturing two-dimensional electrophoresis developed mainly by Schägger and coworkers (Schägger and Von Jagow 1991). Because the first dimension is a native electrophoresis, this approach provides invaluable information upon the assembly of membrane complexes. However, it poses in turn tricky solubilization problems, as the powerful ionic detergents cannot be used because of their denaturing power. The answer to this difficult problem is the combined used of a salt (or a dielectric compound such as aminocaproic acid), a neutral, non-denaturing detergent and a reagent bringing additional electrical charges to the protein complexes to increase their solubility and prevent their aggregation during electrophoresis. This charge-transfer reagent is usually a protein dye such as Coomassie Blue.

Because of the complexity of this solubilization problem, the optimization of the system is usually difficult, and this system has been most often used for mitochondrial and chloroplastic membrane complexes (reviewed in Nijtmans et al. 2002), where it has shown the ability to analyze even very hydrophobic membrane proteins (Devreese et al. 2002). However, this separation approach has been recently applied to whole cell extracts (Camacho-Carjaval et al. 2004) and also to nuclear proteins (Novakova et al. 2006)

### 3.2. Constraints proteomics based on two-dimensional gel electrophoresis

In this scheme, the proteins are first separated by isoelectric focusing (IEF) followed by SDS electrophoresis. The constraints made on sample preparation are thus those induced by the IEF step. One of these constraints is the impossibility to use ionic detergents at high concentrations, as they would mask the protein charge and thus dramatically alter its isoelectric point (pI). Ionic detergents can however be used at low doses to enhance initial solubilization (Wilson et al. 1977), but their amount is limited by the capacity of the IEF system (in terms of ions tolerated) and by the efficiency of the detergent exchange process with takes place during the IEF step. Another major constraint induced by IEF is the requirement for low ionic strength, induced by the high electric fields required for pushing the proteins to their isoelectric points. This means in turn that only uncharged compounds can be used to solubilize proteins, i.e. neutral chaotropes and detergents. The basic solubilization solutions for IEF thus contain high concentrations of a non-ionic chaotrope, historically urea but now more and more a mixture of urea and thiourea (Rabilloud et al. 1997), together with a reducer and a nonionic detergent. While CHAPS and Triton X-100 are the most popular detergents, it has been recently shown that other detergents can enhance the solubility of proteins and give better performances (Chevallet et al. 1998, Luche et al. 2003). In this solubilization process, detergents play a multiple role. They bind to proteins and help to keep them in solution, but they also break protein-lipids interactions and promote lipid solubilization.

Last but certainly not least, it must be recalled that proteins are at their pI at the end of the IEF process, and the pI is the solubility minimum. This means that solubility problems are very important in IEF. this is especially true with native proteins, and IEF is thus best performed with denatured proteins. However, even in this case, many proteins such as membrane proteins are poorly soluble under IEF conditions. Compared with detergent-based electrophoresis, this problem is further enhanced by the fact that only electrically neutral detergents can be used. Ionic detergents would modify the pI of the proteins and thus prevent any correct separation through this parameter. This however means in turn that the beneficial electrostatic repulsion effect cannot be used in isoelectric focusing, thereby leaving proteins under conditions in which their solubility is minimal. While this is not so much a problem for proteins that are normally water-soluble, this turns to be a major problem for proteins that are poorly water-soluble, and especially transmembrane proteins. This explains why this class of proteins is strongly under-represented in this type of separation (for review see Santoni et al. 2000)

### 3.3. Constraints in peptide separation-based approaches

In these methods, the sample preparation process is designed to offer optimal digestion of the proteins into peptides. This means that proteins must be extracted from the sample and denatured to maximize exposure of the protease cleavage sites. This also means that the protease used for peptide production must be active in the separation method. In this case, the robustness of many proteases is a clear advantage. Classical extraction media usually contain either multimolar concentrations of chaotropes or detergents. In the latter case, the sample is usually solubilized and denatured in high concentrations of ionic detergents, and simple dilution is used to bring the detergent concentration down to a point compatible with other steps such as chemical labelling or proteolysis.

The choice between chaotropes and detergent is driven mainly by the constraints imposed by the peptide separation method. In the wide-scope approach based on online two-dimensional chromatography of complex peptide mixtures (Washburn et al., 2001), both the ion exchange and reverse phase steps are very sensitive to detergents. It must be mentioned that these online two-dimensional chromatographic methods are one of the rare cases where the interface between the two separation methods does not bring extra robustness, so that the sample preparation must be compatible with both chromatographic methods. The modification approach (Gevaert et al, 2003), which uses intensively reverse phase chromatography, is also very sensitive to detergent interference. This rules out the use of detergent and favors the use of chatropes, generally nonionic ones because of the ion exchange step. Urea is used for these methods, with possible artefacts induced by urea-driven carbamylation of the sample during the lengthy digestion process. Inclusion of thiourea and lowering of the urea concentration could decrease the incidence of carbamylation in these methods. In methods where a detergent-resistant method is used, e.g. avidin selection of biotinylated peptides (Gygi et al. 1998), extraction by SDS is clearly the method of choice, as the above-mentioned drawbacks are absent.

## 3.4. Constraints in protein chromatography-based approaches

In these methods, the separation is carried out by a chromatographic setup. This setup may use various chromatographic principles for protein separation, and this choice will of course alter the possibilities for protein solubilization. For example, the use of an ion-exchange step (Szponarski et al. 2004) will preclude the use of ionic detergents and thus produce a solubilization situation close to the one observed in 2D gels. However, an important advantage of ion exchange over chromatofocusing or IEF is that the proteins do not need to reach their pI in this separation scheme. This allows in turn to increase protein solubility, which is a major concern in transmembrane protein analysis.

Here again, not all chromatographic setups are usable for any proteomics question. The use of protein reverse phase chromatography, which has been advocated for plasma proteomics (Moritz et al. 2005), precludes in turn the use of any detergent of any type. This prevents the use of this chromatographic setup in most subcellular proteomics experiments, where detergents must be used to solubilize the membrane limiting the subcellular compartments

## 4. Concluding remarks

It should be obvious from the above that membrane proteomics strictly follows Murphy's law. This is due to the fact that there is a mutual exclusion, for physico-chemical reasons, between on the one hand the conditions that must be used to solubilize in water all membrane proteins, including the most hydrophobic ones, and thus give a fair representation of the protein population in the sample, and on the other hand the conditions prevailing in the high resolution peptide separation methods. On top of this problem, there is a second problem linked to the membrane vs. aqueous phase volume in many subcellular preparations, which makes transmembrane proteins rare compared to water-soluble proteins.

Thus, everything concurs in making membrane proteomics probably one of the most difficult sub-field in proteomics. Up to now, the proteomics toolobox that we can use has proven quite imperfect to provide a thorough, quantitative and precise (i.e. including post-translational modifications analysis) analysis of membrane proteins